\documentclass[aps,prl,reprint,amsmath,amssymb,floatfix,superscriptaddress,nofootinbib]{revtex4-1}

\usepackage{amsmath}
\usepackage{amsfonts}
\usepackage{amssymb}
\usepackage{graphicx}			
\usepackage{subfig}				
\usepackage{epstopdf}			
\usepackage{bm}				
\usepackage{units}				
\usepackage{dcolumn}			
\usepackage{verbatim}			
\usepackage[font=footnotesize]{caption}
\usepackage{xcolor}				
\begin{document}

\title{Particle Velocity Distributions in Developing Magnetized Collisionless Shocks in Laser-Produced Plasmas}

\author{D. B. Schaeffer}
	\email{dereks@princeton.edu}
	\affiliation{Department of Astrophysical Sciences,  Princeton University, Princeton, NJ 08540, USA}
\author{W. Fox}
	\affiliation{Department of Astrophysical Sciences,  Princeton University, Princeton, NJ 08540, USA}
   	\affiliation{Princeton Plasma Physics Laboratory, Princeton, New Jersey 08543, USA}
\author{R. K. Follett}
   	\affiliation{Laboratory for Laser Energetics, University of Rochester, Rochester, New York 14623, USA}
\author{G. Fiksel}
   	\affiliation{Center for Ultrafast Optical Science, University of Michigan, Ann Arbor, MI 48109, USA}
\author{C. K. Li}
	\affiliation{Plasma Science and Fusion Center, Massachusetts Institute of Technology, Cambridge, MA, USA}
\author{J. Matteucci}
	\affiliation{Department of Astrophysical Sciences,  Princeton University, Princeton, NJ 08540, USA}
\author{A. Bhattacharjee}
   	\affiliation{Department of Astrophysical Sciences,  Princeton University, Princeton, NJ 08540, USA}
	\affiliation{Princeton Plasma Physics Laboratory, Princeton, New Jersey 08543, USA}
\author{K. Germaschewski}
   	\affiliation{Space Science Center, University of New Hampshire, Durham, New Hampshire 03824, USA}	

\date{\today}

\begin{abstract}

We present the first laboratory observations of time-resolved electron and ion velocity distributions in forming, magnetized collisionless shocks.  Thomson scattering of a probe laser beam was used to observe the interaction of a laser-driven, supersonic piston plasma expanding through a magnetized ambient plasma.  From the Thomson-scattered spectra we measure time-resolved profiles of electron density, temperature, and ion flow speed, as well as spatially-resolved magnetic fields from proton radiography.  We observe direct evidence of the sweeping up and acceleration of ambient ions, magnetic field compression, and the subsequent deformation of the piston ion flow, key steps in shock formation.  Even before the shock has fully formed, we observe strong density compressions and electron heating associated with the pile up of piston ions.  The results demonstrate that laboratory experiments can probe particle velocity distributions relevant to collisionless shocks, and can complement, and in some cases overcome, the limitations of similar measurements undertaken by spacecraft missions.

\end{abstract}

\maketitle


Collisionless shocks are commonly found in systems in which strongly-driven flows interact with pre-existing magnetic fields, including planetary bow shocks in the heliosphere \cite{smith_jupiters_1975,smith_saturns_1980,sulaiman_quasiperpendicular_2015} and astrophysical shocks in supernova remnants \cite{spicer_model_1990,bamba_small-scale_2003}.  In collisionless plasmas, these shocks form on spatial scales much smaller than the collisional mean free path due to dissipation mediated by electromagnetic fields.  For most observed shocks, the fast inflow of particles can only be managed through the magnetic reflection of some particles back upstream, resulting in complex interactions between populations of inflowing, reflected, and shocked ions and electrons that are not fully understood.  Consequently, fundamental questions, such as how energy is partitioned between electrons and ions across a collisionless shock \cite{balikhin_new-mechanism_1993,lembege_demagnetization_2003,schwartz_electron_2011}, remain unanswered.

A key method for addressing these questions is the direct probing of particle velocity distributions, which has primarily been undertaken through \textit{in situ} measurements by spacecraft.  These missions have yielded a wealth of information on shock physics \cite{burgess_collisionless_2015}, and have recently begun to address the question of energy partitioning \cite{chen_electron_2018} as improved diagnostics have allowed high-resolution sampling of velocity distributions.  Even so, spacecraft remain fundamentally limited, as they rely on the inherently noisy process of sampling shock crossings through multiple orbits and have difficulty gauging large-scale, 3D effects due to undersampling \cite{lobzin_nonstationarity_2007,johlander_rippled_2016}.  Laboratory experiments, with reproducible and controllable plasma conditions, can complement and overcome some of these limitations to help address fundamental questions \cite{howes_laboratory_2018}, and have recently extended the regimes of magnetized shock formation to strongly-driven laser plasmas \cite{niemann_observation_2014,schaeffer_generation_2017}.  Moreover, velocity distributions can be similarly probed in the laboratory by measuring the Thomson scattering of light off plasma waves \cite{froula_thomson-scattering_2006,sheffield_plasma_2011}.  Early experiments \cite{paul_measurement_1967,desilva_observation_1967} pioneered the use of Thomson scattering to study magnetized shocks, but were limited to a sparse sampling of the electron velocity distribution.  Recent experiments have used this diagnostic to study velocity distributions in collisional shocks \cite{suttle_structure_2016,rinderknecht_highly_2018} and in unmagnetized collisionless counter-streaming flows \cite{ross_collisionless_2013}.

In this Letter, we present the first laboratory observations of temporally-resolved electron and ion velocity distributions in forming, magnetized collisionless shocks.  The distributions were acquired through Thomson scattering of a probe laser that diagnosed the interaction of a laser-driven, supersonic piston plasma expanding through a magnetized ambient plasma.  Spatially-resolved 2D proton radiography images of the magnetic field were also acquired.  We directly observe the interplay between the piston and ambient plasmas in the initial stages of shock formation, including the acceleration of ambient ions and the pile up of piston ions behind the resulting compressed magnetic field.  These effects are found to depend critically on the density of the ambient plasma and the presence of the background magnetic field.  The results build on an experimental platform that has studied high-Mach-number magnetized collisionless shocks \cite{schaeffer_generation_2017,schaeffer_high-mach_2017}, laser-driven magnetic reconnection \cite{fiksel_magnetic_2014}, and Weibel-mediated shocks \cite{fox_filamentation_2013}.


\begin{figure}[t]
	\centering
	\includegraphics{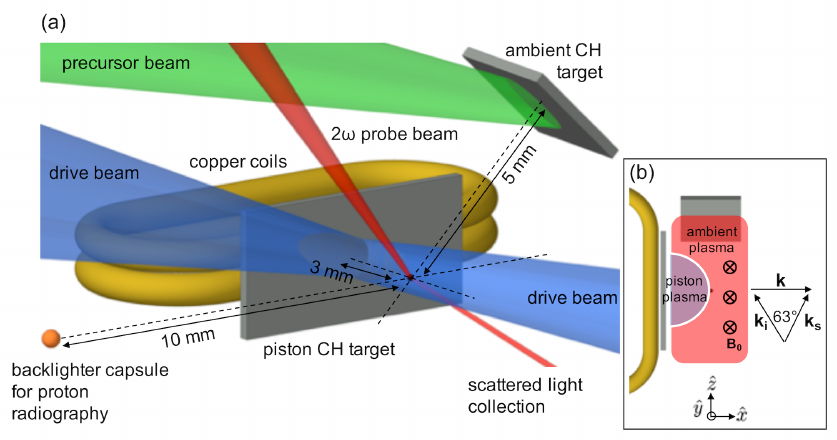} 
	\caption{(a) Experimental setup.  A background magnetic field primarily directed along $\hat{y}$ is pre-imposed using current-carrying copper wires.  A precursor laser ablates a CH target to create a magnetized ambient plasma.  Two drive beams then generate a CH piston plasma that expands through the ambient plasma to drive a shock.  Temperature, density, and velocity are diagnosed in the $\hat{x}$ direction using Thomson scattering with a 2$\omega$ probe beam.  20 beams (not shown) compress a DHe3 backlighter capsule to generate mono-energetic protons that probe the magnetic field structure in the $x$-$y$ plane. (b) Top-down schematic view of the setup and Thomson scattering geometry.}
	\label{fig:setup}
\end{figure}

\begin{figure*}[t]
	\centering
	\includegraphics{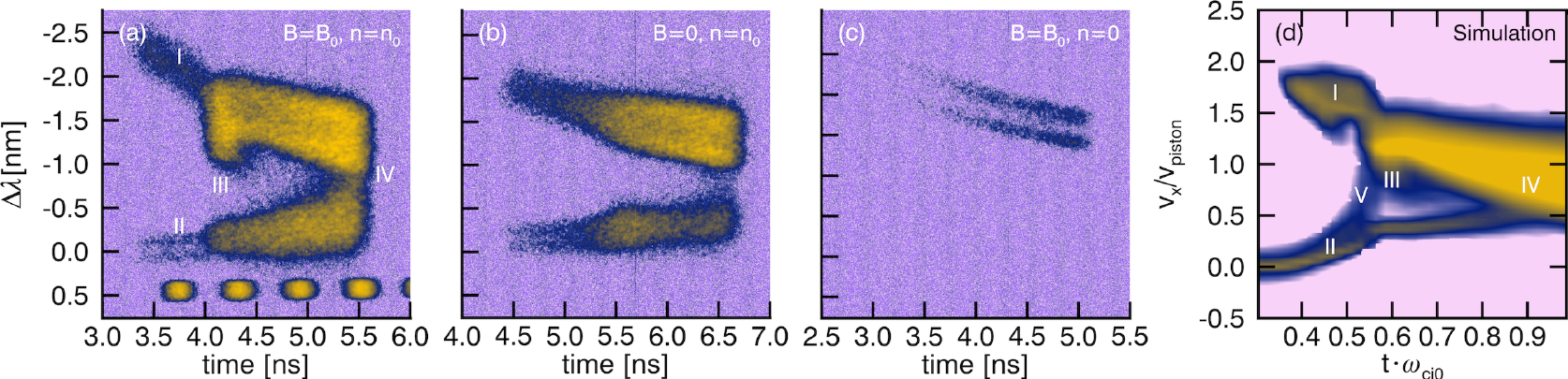} 
	\caption{IAW spectra of piston-ambient interactions under three experimental conditions: (a) magnetized ambient plasma, (b) unmagnetized ambient plasma, and (c) no ambient plasma.  Data in (a) and (c) was taken at $x=3$ mm (TCC), while (b) was taken at $x=4$ mm. The marks at the bottom of (a) are timing fiducials.  (d) Simulated ion velocity space in conditions similar to (a), with velocity relative to thepiston speed and time relative to the upstream gyrofrequency.  Regions of interest are labeled with Roman numerals.}
	\label{fig:iaw}
\end{figure*}

\begin{figure}[t]
	\centering
	\includegraphics{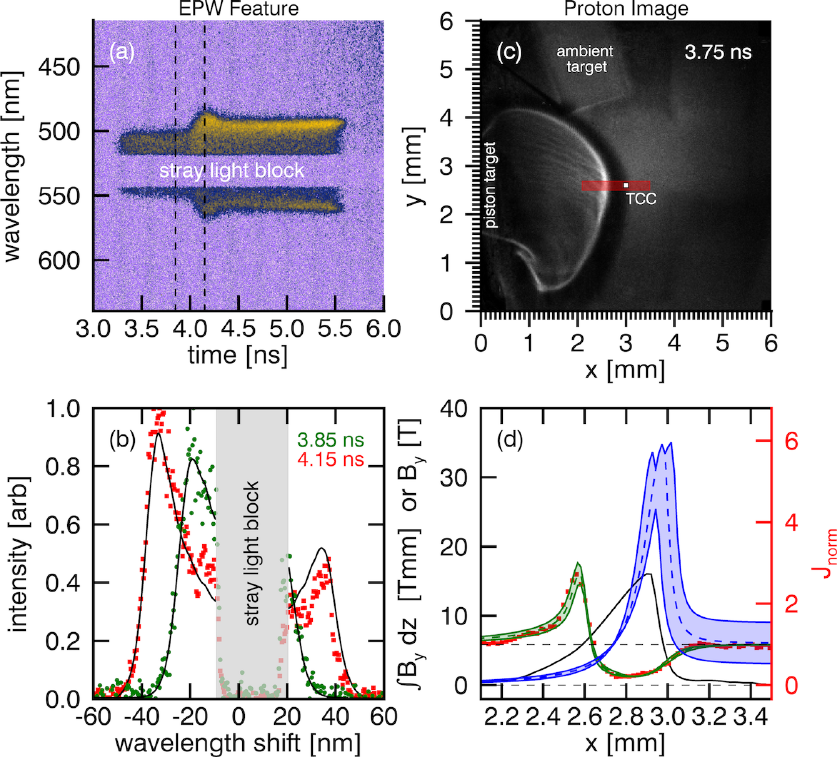} 
	\caption{(a) Streaked Thomson-scattered spectrum of the EPW feature taken at TCC, corresponding to Fig.~\ref{fig:iaw}a.  (b) Two example profiles at time $t=t_0+3.85$ ns (green) and $t=t_0+4.15$ ns (red), along with best fits (black). (c) Proton radiography image taken at time $t=t_0+3.75$ ns using 14.7 MeV protons.  (d) Proton intensity (red squares) taken from the red region in (c), normalized to the mean intensity, and the associated reconstructed path-integrated magnetic field $\int B_y dz$ (black).  Also shown is the normalized proton intensity (green dashed) forwarded-modeled from a 2D synthetic magnetic field $B_y(x,z)$, which has the dashed blue profile at $z=0$.  The model uncertainties are shown as shaded regions.}
	\label{fig:img_sum}
\end{figure}

\begin{figure*}[t]
	\centering
	\includegraphics{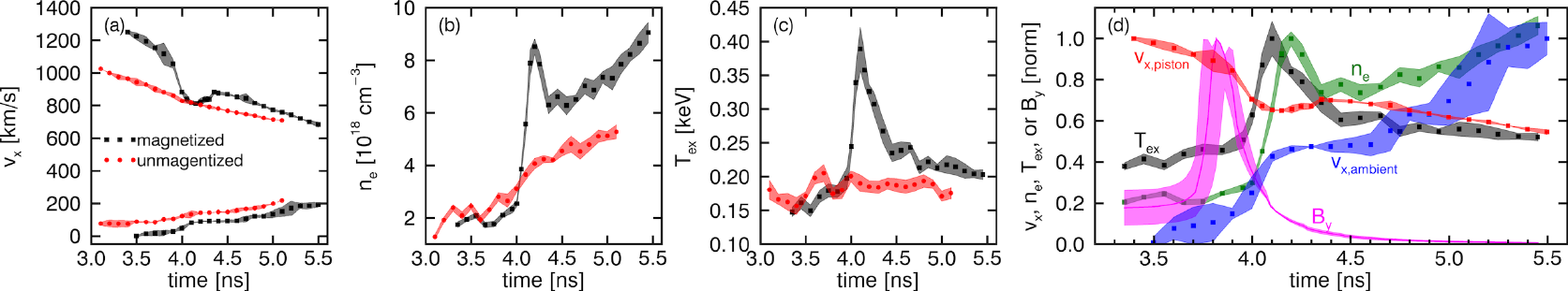} 
	\caption{Thomson scattering results. Measured (a) piston and ambient ion flow speed $v_x$, (b) electron density $n_e$, and (c) electron temperature $T_{ex}$ for a piston plasma expanding through a magnetized (black squares) and unmagnetized (red circles) ambient plasma.  (d) Electron temperature (black), density (green), magnetic field (magenta), and piston (red) and ambient (blue) ion flow speed for the magnetized case.  The magnetic field profile was constructed from the model in Fig.~\ref{fig:img_sum}d.  Error bars are shown as shaded regions.  The magnetized plasma were probed at $x=3$ mm (TCC), while the unmagnetized plasma was probed at $x=4$ mm.  The unmagnetized data has been shifted forward in time by 1.5 ns for ease of comparison.}
	\label{fig:ts_data}
\end{figure*}


\textit{Setup.} The experiments were carried out on the OMEGA laser facility \cite{boehly_the-upgrade_1995} and are shown schematically in Fig. \ref{fig:setup}.  The experiment utilizes two planar CH targets and a set of copper coils to generate a magnetic field.  The ``piston'' target is attached to the coils 3 mm from target chamber center (TCC) and defines the experimental coordinate system, with $\hat{x}$ along the target normal, $\hat{y}$ parallel to the long edge, and $\hat{z}$ parallel to the short edge.  A second ``ambient'' target is centered at TCC along $\hat{x}$ and offset 5 mm diagonally at a 45$^{\circ}$ angle.  A background magnetic field is generated by the coils \cite{fiksel_note:_2015}.  The initial field $B_y$ has a peak strength of 10 T near the piston target and falls off like $1/x$ along $\hat{x}$, while it is nearly uniform across the target surface due to the elongated and stacked coil structure.  A precursor beam (1053 nm, 100 J, 1 ns) incident on the ambient target creates a plasma that expands through the background field.  As shown previously \cite{fiksel_magnetic_2014,schaeffer_high-mach_2017}, over 12 ns this plasma fills the volume in front of the piston target and mixes with the background field to create a magnetized ambient plasma.  Two drive beams (1053 nm, 350 J, 2 ns) incident on the piston target at time $t_0$ then generate a supersonic piston plasma, which expands through the ambient plasma.

The primary diagnostic was temporally-resolved Thomson scattering using a 2$\omega$ probe beam (527 nm, 30-50 J, 2 ns) \cite{follett_plasma_2016}.  Scattered light from the probe beam was collected from a localized volume ($50\times50\times70$ $\mu$m$^{3}$) such that the probed wavevector $\bf{k}=\bf{k_i}-\bf{k_s}$ was directed along the piston expansion direction (\textit{i.e.} along $\hat{x}$), where $\bf{k_i}$ is the incident wavevector and $\bf{k_s}$ is the scattered wavevector (Fig.~\ref{fig:setup}b).  The scattering angle was 63$^{\circ}$, yielding a scattering parameter $\alpha=1/k\lambda_{de}\approx 1.5$ for typical plasma parameters and placing the scattered signal in the collective regime.  The collected light was split along two beam paths.  One path measured light scattered from electron plasma waves (EPW), which can provide information on the electron density and temperature.  The other path measured light scattered from ion acoustic waves (IAW), which can also diagnose the electron temperature, as well as the ion temperature and flow speed.  The EPW and IAW signals were passed through spectrometers with wavelength resolutions of 0.5 and 0.05 nm, respectively, and imaged onto streak cameras with a temporal resolution of 50 ps.  The location of the probed plasma ranged from 3 to 4 mm from the piston target along $\hat{x}$.  The scattered signal was streaked for 2 ns starting 3 to 4.5 ns after $t_0$.

The magnetic field structure was measured using proton radiography \cite{petrasso_lorentz_2009}.  A 420 $\mu$m diameter glass capsule filled with DHe3 was placed 10 mm from TCC along $\hat{z}$ and irradiated by 20 beams at $t_0+3$ ns.  The resulting implosion produced 3 and 14.7 MeV protons as fusion by-products, which passed through the plasma and were collected on CR-39 plates placed 154 mm from TCC (geometric magnification $M=16.4$).  The protons leave tracks in the CR-39 that correspond to a 2D map of proton deflections in the $x$-$y$ plane, which can be converted to path-integrated magnetic field amplitudes.


\textit{Results.} Fig. \ref{fig:iaw} shows streaked IAW spectra taken under three experimental configurations: (a) a magnetized piston-ambient interaction, (b) an unmagnetized piston-ambient interaction, and (c) a magnetized piston expansion.  The EPW spectrum corresponding to Fig. \ref{fig:iaw}a is shown in Fig.~\ref{fig:img_sum}a, and a proton radiograph taken under the same conditions is shown in Fig. \ref{fig:img_sum}c.  The ambient plasma was measured at TCC using Thomson scattering in the absence of a piston plasma over the same time intervals as in Fig.~\ref{fig:iaw}.  The measurements yielded a time-averaged mean electron density $n_{e0}=0.9\pm 0.2\times10^{18}$ cm$^{-3}$ and temperature $T_{e0}=40\pm 10$ eV \cite{Note1}.

The spectra show qualitative signatures of a developing magnetized collisionless shock, and can be divided into four distinct regions in the IAW spectra, labeled I-IV in Fig. \ref{fig:iaw}a.  Region I consists of piston ions that are streaming through the ambient plasma (region II) but largely unaffected by the magnetic field.  A key step in piston-driven shock formation is the sweeping up of ambient plasma \cite{bondarenko_collisionless_2017} and the resulting compression of the magnetic field.  The increased field then causes a pile up of piston plasma and deformation of the piston flow.  Both the ambient ion acceleration and piston deformation are seen in region III, which also corresponds to the peak in the EPW spectra in Fig.~\ref{fig:img_sum}a.  Eventually, most of the ambient ions not participating in shock formation are swept up by the piston, which results in the merging of the piston and ambient plasmas in region IV.  Without a background magnetic field (Fig. \ref{fig:iaw}b), no ion pile up or flow deformation is observed, though the ambient ions are still eventually swept up.  Likewise, Fig. \ref{fig:iaw}c shows that with only a magnetized piston plasma, no shock forms.  These last two cases indicate that the presence of both the ambient plasma and background field is critical to shock formation.  Lastly, Fig.~\ref{fig:iaw}d is the $x$ component of the ion velocity distribution in the Thomson-scattering volume as a function of time from a 1D \textsc{psc} \cite{germaschewski_the-plasma_2016,fox_kinetic_2018} particle-in-cell simulation under conditions similar to Fig.~\ref{fig:iaw}a.  The four regions of Fig.~\ref{fig:iaw}a are clearly visible in the simulation and show that there is strong correspondence between the velocity distributions and the Thomson-scattered spectra.  There is an additional intriguing feature in region V: the formation of a shock in the ambient H plasma just ahead of the piston pile up.  We do not directly observe H shock features in the spectra, though calculations indicate that the H ion acoustic waves would be heavily Landau damped relative to the C waves.  

Fig.~\ref{fig:img_sum}c shows a 14.7 MeV proton image taken at $t_0+3.75$ ns under the same conditions as Fig.~\ref{fig:iaw}a.  The magnetic cavity created by the piston can be clearly seen outlined by white, high-proton-fluence and dark, low-proton-fluence ribbons that result from the deflection of protons by the $B_y$ magnetic field.  The variation between dark and light fluence represents a large gradient in (path-integrated) magnetic field strength associated with the forming shock.  This can be seen in Fig.~\ref{fig:img_sum}d, where we reconstruct the line-integrated magnetic field $\int B_y dz$ (black line) along a 1D profile through TCC by inverting the corresponding proton fluence profile (red squares).  To unfold the original field, we assume a form for $B_y(x,z)$ and forward model a synthetic proton fluence.  By optimizing the parameters of the model, we find good agreement between the data and the synthetic proton fluence (green line).  At $z=0$, corresponding to the location of the Thomson scattering measurements, the model field $B_y(x,0)$ (blue line) has a peak value $B_{y,peak}=35\pm3$ T at $x_{peak}=2.98\pm0.05$ mm, though the upstream value $B_{y0}=6\pm3$ T is not well constrained.  Here, the uncertainties are derived by comparing best fits at different fixed upstream values.  Similar results are obtained from the 3 MeV proton image, indicating that the protons are primarily deflected by magnetic fields rather than electric fields.

We can further quantify the Thomson-scattered spectra in Figs.~\ref{fig:iaw} \& \ref{fig:img_sum} by iteratively fitting the data with a spectral model of the scattered power \cite{follett_plasma_2016}.  Time-resolved parameters can then be extracted, including electron density $n_e$ and the $x$-component of the electron temperature $T_{ex}$ and ion flow speed $v_x$.  An example EPW spectrum and fit is shown in Fig. \ref{fig:img_sum}b.  To perform error analysis, we employ a Monte Carlo approach in which the extracted plasma parameters represent the mean value over 50 fits, with error bars corresponding to the standard deviation.  In all cases, the EPW spectral fits assumed Maxwellian velocity distributions.  In contrast, the IAW spectra involve multiple ion species (C and H) and multiple flows from potentially non-Maxwellian ion distributions.  Extracting parameters from these spectra is beyond the scope of this Letter and will be reported separately.  Instead, we only determine the ion flow speed from the Doppler shifts of the spectra, which can be accurately resolved without knowing the exact form of the scattered power \cite{sheffield_plasma_2011}.  Based on the results of these fits (see Fig.~\ref{fig:ts_data}), we can justify the use of Maxwellian distributions by estimating the electron $\tau_{ee}$ and electron-ion $\tau_{ei}$ collision times relative to the fastest gradient timescales $\tau_{s}\sim200$ ps and the electron plasma frequency $\omega_{pe}$.  We find for the electrons that $\tau_{ee}<\tau_s$, indicating that the electrons are well thermalized, and that $\tau_{pe}\ll\tau_{ei}<\tau_{ee}$, so that collisions do not significantly affect the EPW spectra.  Furthermore, the shock layer is dominantly determined by the piston-ambient ion (and eventually ambient-ambient ion) interaction, which is highly collisionless ($\tau_{pa}/\tau_{s}\ggg1$) due the the large flow velocities in these experiments.


A summary of the Thomson scattering results is shown in Fig.~\ref{fig:ts_data} for the two piston-ambient interactions: magnetized (black) and unmagnetized (red).  Fig.~\ref{fig:ts_data}a shows two sets of flow speeds $v_x$ extracted from the IAW spectra and corresponding to the faster (piston) and slower (ambient) moving populations.  For the magnetized case, the piston ions exhibit a rapid deceleration around $t_0+4.0$ ns, coincident with the onset of the region of ion pile up in Fig. \ref{fig:iaw}a.  Over the same time the ambient ions are accelerated, and then plateau for several hundred ps before being accelerated again as they begin to merge with the piston plasma.  In the unmagnetized case, the piston ions show no deceleration and are consistent with a free-streaming expansion ($v\propto1/t$).  Figs.~\ref{fig:ts_data}b-c show electron density and temperature extracted from the EPW spectra.  In the region of ion pile up, the magnetized case exhibits a strong density compression $n_e/n_{e0}\approx10$, steep density ramp $\tau_{n}\sim200$ ps, and electron heating $T_{e\perp}/T_{e0}\approx 10$, indicating that these effects are necessary but not sufficient signatures of shock formation.  No density compression or electron heating is observed in the unmagnetized case.

Fig.~\ref{fig:ts_data}d combines temperature (black), density (green), magnetic field (purple), and piston (red) and ambient (blue) ion flow results for the magnetized case.  The field is plotted assuming that it is slowly changing on the timescales of interest, so that the spatial profile can be converted to a temporal profile using the time-of-flight speed $v_{field}=790\pm20$ km/s.  The combined profiles paint a self-consistent picture of the initial stages of shock formation.  Firstly, the magnetic field acts as an interface between the highly-magnetized ambient and piston electrons: swept-over ambient electrons compress the field at the leading edge while piston electrons expel the field \cite{wright_early-time_1971}.  Consequently, the piston electrons (and ions) will necessarily pile up behind the magnetic compression, as observed.  This results in a localized electron density peak that then transitions into the smooth ablation profile of the piston plume.  The temperature in turn rises adiabatically ($T_e\propto n_e^{2/3}$) with the density, consistent with collisional electrons.  

While at this stage in formation the density profile primarily reflects piston dynamics, it also crucially leads to the sweeping up of ambient ions through the pressure gradient electric field $E_x=\nabla P_e/e n _e$, where $P_e = n_e T_e$.  This is directly observed in Fig.~\ref{fig:ts_data}d, where the change in ambient ion speed between 3.9 and 4.15 ns ($\Delta v\sim40$ km/s) is quantitatively consistent with an acceleration due to $E_x$ ($\Delta v_E=\int (Z_C e/m_C) E_x dt\sim50$ km/s), assuming that $\nabla P_e\approx (1/v_{field}) dP_e/dt$. After they are accelerated, the ambient ions pass through the developing shock into the proto-downstream region, where they coast until being swept up by the main piston plume. The pressure gradient electric field also accounts for the behavior of the piston ion flow.  Behind the density compression the pressure-gradient field points back towards the main plume, so incoming piston ions are decelerated (seen around 4.3 ns).  Those ions are then strongly accelerated by the oppositely-directed field at the leading edge (the same field as for the ambient ions, though because the piston ions are moving with the density compression, they experience the acceleration for longer and thus obtain a larger change in speed).  The deformation of the piston ion flow is therefore a key signature of the onset of piston-driven shock formation.

In summary, we have measured for the first time through Thomson scattering the evolution of electron and ion velocity distributions of a forming, magnetized collisionless shock.  We have extracted time-resolved profiles of electron temperature, density, and ion flow speed, which indicate the development of strong density compressions and electron heating associated with the pile up of piston ions and acceleration of ambient ions.  Proton radiography images confirm that there is an associated strong magnetic compression in the same region.  This acceleration of ambient ions and subsequent deformation of the piston ion flow is a key component of magnetized shock formation, and is not observed without both a background magnetic field and ambient plasma.  Since the distributions can in principle be probed along any direction, these results will enable future experiments to study multi-dimensional distribution functions in a manner analogous to spacecraft, allowing direct comparisons between studies of space and laboratory collisionless shocks.


\begin{acknowledgments}
We thank the staff of the Omega facility for their help in executing these experiments.  Time on the Omega facility was funded by the Department of Energy (DOE) through grant No. DE-NA0003613.  Processing of the proton images was funded by the DOE under grant No. DE-FG03-09NA29553.  Simulations were conducted on the Titan supercomputer at the Oak Ridge Leadership Computing Facility at the Oak Ridge National Laboratory through the Innovative and Novel Computational Impact on Theory and Experiment (INCITE) program, which is supported by the Office of Science of the DOE under contract No. DE-AC05-00OR22725.  Development of the \textsc{psc} code was funded by the DOE through grant No. DE-SC0008655.  This research was also supported by the DOE under grant No. DE-SC0016249.
\end{acknowledgments}

\bibliographystyle{naturemag.bst}
\bibliography{shock_vel_dist_bib}

\begin{thebibliography}{10}
\expandafter\ifx\csname url\endcsname\relax
  \def\url#1{\texttt{#1}}\fi
\expandafter\ifx\csname urlprefix\endcsname\relax\def\urlprefix{URL }\fi
\providecommand{\bibinfo}[2]{#2}
\providecommand{\eprint}[2][]{\url{#2}}

\bibitem{smith_jupiters_1975}
\bibinfo{author}{Smith, E.~J.} \emph{et~al.}
\newblock \bibinfo{title}{Jupiter's magnetic field. magnetosphere, and
  interaction with the solar wind: Pioneer 11}.
\newblock \emph{\bibinfo{journal}{Science}} \textbf{\bibinfo{volume}{188}},
  \bibinfo{pages}{451--455} (\bibinfo{year}{1975}).

\bibitem{smith_saturns_1980}
\bibinfo{author}{Smith, E.~J.} \emph{et~al.}
\newblock \bibinfo{title}{Saturn's magnetic field and magnetosphere}.
\newblock \emph{\bibinfo{journal}{Science}} \textbf{\bibinfo{volume}{207}},
  \bibinfo{pages}{407--410} (\bibinfo{year}{1980}).

\bibitem{sulaiman_quasiperpendicular_2015}
\bibinfo{author}{Sulaiman, A.~H.} \emph{et~al.}
\newblock \bibinfo{title}{Quasiperpendicular high mach number shocks}.
\newblock \emph{\bibinfo{journal}{Phys. Rev. Lett.}}
  \textbf{\bibinfo{volume}{115}}, \bibinfo{pages}{125001}
  (\bibinfo{year}{2015}).

\bibitem{spicer_model_1990}
\bibinfo{author}{Spicer, D.~S.}, \bibinfo{author}{Maran, S.~P.} \&
  \bibinfo{author}{Clark, R.~W.}
\newblock \bibinfo{title}{A model of the pre-sedov expansion phase of supernova
  remnant-ambient plasma coupling and x-ray emission from sn 1987a}.
\newblock \emph{\bibinfo{journal}{Astrophys. J.}}
  \textbf{\bibinfo{volume}{356}}, \bibinfo{pages}{549--571}
  (\bibinfo{year}{1990}).

\bibitem{bamba_small-scale_2003}
\bibinfo{author}{Bamba, A.}, \bibinfo{author}{Yamazaki, R.},
  \bibinfo{author}{Ueno, M.} \& \bibinfo{author}{Koyama, K.}
\newblock \bibinfo{title}{{Small-Scale} structure of the 1006 shock with
  chandra observations}.
\newblock \emph{\bibinfo{journal}{Astrophys. J.}}
  \textbf{\bibinfo{volume}{589}}, \bibinfo{pages}{827--837}
  (\bibinfo{year}{2003}).

\bibitem{balikhin_new-mechanism_1993}
\bibinfo{author}{Balikhin, M.}, \bibinfo{author}{Gedalin, M.} \&
  \bibinfo{author}{Petrukovich, A.}
\newblock \bibinfo{title}{New mechanism for electron heating in shocks}.
\newblock \emph{\bibinfo{journal}{Phys. Rev. Lett.}}
  \textbf{\bibinfo{volume}{70}}, \bibinfo{pages}{1259--1262}
  (\bibinfo{year}{1993}).

\bibitem{lembege_demagnetization_2003}
\bibinfo{author}{Lemb{\`e}ge, B.}, \bibinfo{author}{Savoini, P.},
  \bibinfo{author}{Balikhin, M.}, \bibinfo{author}{Walker, S.} \&
  \bibinfo{author}{Krasnoselskikh, V.}
\newblock \bibinfo{title}{Demagnetization of transmitted electrons through a
  quasi‐perpendicular collisionless shock}.
\newblock \emph{\bibinfo{journal}{J. Geophys. Res.}}
  \textbf{\bibinfo{volume}{108}} (\bibinfo{year}{2003}).

\bibitem{schwartz_electron_2011}
\bibinfo{author}{Schwartz, S.~J.}, \bibinfo{author}{Henley, E.},
  \bibinfo{author}{Mitchell, J.} \& \bibinfo{author}{Krasnoselskikh, V.}
\newblock \bibinfo{title}{Electron temperature gradient scale at collisionless
  shocks}.
\newblock \emph{\bibinfo{journal}{Phys. Rev. Lett.}}
  \textbf{\bibinfo{volume}{107}}, \bibinfo{pages}{215002}
  (\bibinfo{year}{2011}).

\bibitem{burgess_collisionless_2015}
\bibinfo{author}{Burgess, D.} \& \bibinfo{author}{Scholer, M.}
\newblock \emph{\bibinfo{title}{Collisionless Shocks in Space Plasmas:
  Structure and Accelerated Particles}}.
\newblock Cambridge Atmospheric and Space Science Series
  (\bibinfo{publisher}{Cambridge University Press}, \bibinfo{year}{2015}).

\bibitem{chen_electron_2018}
\bibinfo{author}{Chen, L.-J.} \emph{et~al.}
\newblock \bibinfo{title}{Electron bulk acceleration and thermalization at
  earth's quasiperpendicular bow shock}.
\newblock \emph{\bibinfo{journal}{Phys. Rev. Lett.}}
  \textbf{\bibinfo{volume}{120}}, \bibinfo{pages}{225101}
  (\bibinfo{year}{2018}).

\bibitem{lobzin_nonstationarity_2007}
\bibinfo{author}{Lobzin, V.~V.} \emph{et~al.}
\newblock \bibinfo{title}{Nonstationarity and reformation of high-mach-number
  quasiperpendicular shocks: Cluster observations}.
\newblock \emph{\bibinfo{journal}{Geophys. Res. Lett.}}
  \textbf{\bibinfo{volume}{34}}, \bibinfo{pages}{L05107}
  (\bibinfo{year}{2007}).

\bibitem{johlander_rippled_2016}
\bibinfo{author}{Johlander, A.} \emph{et~al.}
\newblock \bibinfo{title}{Rippled quasiperpendicular shock observed by the
  magnetospheric multiscale spacecraft}.
\newblock \emph{\bibinfo{journal}{Phys. Rev. Lett.}}
  \textbf{\bibinfo{volume}{117}}, \bibinfo{pages}{165101}
  (\bibinfo{year}{2016}).

\bibitem{howes_laboratory_2018}
\bibinfo{author}{Howes, G.~G.}
\newblock \bibinfo{title}{Laboratory space physics: Investigating the physics
  of space plasmas in the laboratory}.
\newblock \emph{\bibinfo{journal}{Phys. Plasmas}}
  \textbf{\bibinfo{volume}{25}}, \bibinfo{pages}{055501}
  (\bibinfo{year}{2018}).

\bibitem{niemann_observation_2014}
\bibinfo{author}{Niemann, C.} \emph{et~al.}
\newblock \bibinfo{title}{Observation of collisionless shocks in a large
  current-free laboratory plasma}.
\newblock \emph{\bibinfo{journal}{Geophys. Res. Lett.}}
  \textbf{\bibinfo{volume}{41}}, \bibinfo{pages}{7413--7418}
  (\bibinfo{year}{2014}).

\bibitem{schaeffer_generation_2017}
\bibinfo{author}{Schaeffer, D.~B.} \emph{et~al.}
\newblock \bibinfo{title}{Generation and evolution of high-mach-number
  laser-driven magnetized collisionless shocks in the laboratory}.
\newblock \emph{\bibinfo{journal}{Phys. Rev. Lett.}}
  \textbf{\bibinfo{volume}{119}}, \bibinfo{pages}{025001}
  (\bibinfo{year}{2017}).

\bibitem{froula_thomson-scattering_2006}
\bibinfo{author}{Froula, D.~H.}, \bibinfo{author}{Ross, J.~S.},
  \bibinfo{author}{Divol, L.} \& \bibinfo{author}{Glenzer, S.~H.}
\newblock \bibinfo{title}{Thomson-scattering techniques to diagnose local
  electron and ion temperatures, density, and plasma wave amplitudes in laser
  produced plasmas (invited)}.
\newblock \emph{\bibinfo{journal}{Rev. Sci. Instrum.}}
  \textbf{\bibinfo{volume}{77}}, \bibinfo{pages}{10E522}
  (\bibinfo{year}{2006}).

\bibitem{sheffield_plasma_2011}
\bibinfo{author}{Sheffield, J.}, \bibinfo{author}{Froula, D.~H.},
  \bibinfo{author}{Glenzer, S.~H.} \& \bibinfo{author}{Luhmann, N.}
\newblock \emph{\bibinfo{title}{Plasma Scattering of Electromagnetic
  Radiaiton}} (\bibinfo{publisher}{Academic Press}, \bibinfo{year}{2011}),
  \bibinfo{edition}{2nd} edn.

\bibitem{paul_measurement_1967}
\bibinfo{author}{Paul, J. W.~M.}, \bibinfo{author}{Goldenbaum, G.~C.},
  \bibinfo{author}{Iiyoshi, A.}, \bibinfo{author}{Holmes, L.~S.} \&
  \bibinfo{author}{Hardcastle, R.~A.}
\newblock \bibinfo{title}{Measurement of electron temperatures produced by
  collisionless shock waves in a magnetized plasma}.
\newblock \emph{\bibinfo{journal}{Nature}} \textbf{\bibinfo{volume}{216}},
  \bibinfo{pages}{363--364} (\bibinfo{year}{1967}).

\bibitem{desilva_observation_1967}
\bibinfo{author}{{DeSilva}, A.~W.} \& \bibinfo{author}{Stamper, J.~A.}
\newblock \bibinfo{title}{Observation of anomalous electron heating in plasma
  shock waves}.
\newblock \emph{\bibinfo{journal}{Phys. Rev. Lett.}}
  \textbf{\bibinfo{volume}{19}}, \bibinfo{pages}{1027} (\bibinfo{year}{1967}).

\bibitem{suttle_structure_2016}
\bibinfo{author}{Suttle, L.~G.} \emph{et~al.}
\newblock \bibinfo{title}{Structure of a magnetic flux annihilation layer
  formed by the collision of supersonic, magnetized plasma flows}.
\newblock \emph{\bibinfo{journal}{Phys. Rev. Lett.}}
  \textbf{\bibinfo{volume}{116}}, \bibinfo{pages}{225001}
  (\bibinfo{year}{2016}).

\bibitem{rinderknecht_highly_2018}
\bibinfo{author}{Rinderknecht, H.~G.} \emph{et~al.}
\newblock \bibinfo{title}{Highly resolved measurements of a developing strong
  collisional plasma shock}.
\newblock \emph{\bibinfo{journal}{Phys. Rev. Lett.}}
  \textbf{\bibinfo{volume}{120}}, \bibinfo{pages}{095001}
  (\bibinfo{year}{2018}).

\bibitem{ross_collisionless_2013}
\bibinfo{author}{Ross, J.~S.} \emph{et~al.}
\newblock \bibinfo{title}{Collisionless coupling of ion and electron
  temperatures in counterstreaming plasma flows}.
\newblock \emph{\bibinfo{journal}{Phys. Rev. Lett.}}
  \textbf{\bibinfo{volume}{110}}, \bibinfo{pages}{145005}
  (\bibinfo{year}{2013}).

\bibitem{schaeffer_high-mach_2017}
\bibinfo{author}{Schaeffer, D.~B.} \emph{et~al.}
\newblock \bibinfo{title}{High-mach number, laser-driven magnetized
  collisionless shocks}.
\newblock \emph{\bibinfo{journal}{Phys. Plasmas}}
  \textbf{\bibinfo{volume}{24}}, \bibinfo{pages}{122702}
  (\bibinfo{year}{2017}).

\bibitem{fiksel_magnetic_2014}
\bibinfo{author}{Fiksel, G.} \emph{et~al.}
\newblock \bibinfo{title}{Magnetic reconnection between colliding magnetized
  laser-produced plasma plumes}.
\newblock \emph{\bibinfo{journal}{Phys. Rev. Lett.}}
  \textbf{\bibinfo{volume}{113}}, \bibinfo{pages}{105003}
  (\bibinfo{year}{2014}).

\bibitem{fox_filamentation_2013}
\bibinfo{author}{Fox, W.} \emph{et~al.}
\newblock \bibinfo{title}{Filamentation instability of counterstreaming
  laser-driven plasmas}.
\newblock \emph{\bibinfo{journal}{Phys. Rev. Lett.}}
  \textbf{\bibinfo{volume}{111}}, \bibinfo{pages}{225002}
  (\bibinfo{year}{2013}).

\bibitem{boehly_the-upgrade_1995}
\bibinfo{author}{Boehly, T.~R.} \emph{et~al.}
\newblock \bibinfo{title}{The upgrade to the omega laser system}.
\newblock \emph{\bibinfo{journal}{Rev. Sci. Instrum.}}
  \textbf{\bibinfo{volume}{66}}, \bibinfo{pages}{508--510}
  (\bibinfo{year}{1995}).

\bibitem{fiksel_note:_2015}
\bibinfo{author}{Fiksel, G.} \emph{et~al.}
\newblock \bibinfo{title}{Note: Experimental platform for magnetized
  high-energy-density plasma studies at the omega laser facility}.
\newblock \emph{\bibinfo{journal}{Rev. Sci. Instrum.}}
  \textbf{\bibinfo{volume}{86}} (\bibinfo{year}{2015}).

\bibitem{follett_plasma_2016}
\bibinfo{author}{Follett, R.~K.} \emph{et~al.}
\newblock \bibinfo{title}{Plasma characterization using ultraviolet thomson
  scattering from ion-acoustic and electron plasma waves (invited)}.
\newblock \emph{\bibinfo{journal}{Rev. Sci. Instrum.}}
  \textbf{\bibinfo{volume}{87}}, \bibinfo{pages}{11E401}
  (\bibinfo{year}{2016}).

\bibitem{petrasso_lorentz_2009}
\bibinfo{author}{Petrasso, R.~D.} \emph{et~al.}
\newblock \bibinfo{title}{Lorentz mapping of magnetic fields in hot dense
  plasmas}.
\newblock \emph{\bibinfo{journal}{Phys. Rev. Lett.}}
  \textbf{\bibinfo{volume}{103}}, \bibinfo{pages}{085001}
  (\bibinfo{year}{2009}).

\bibitem{Note1}
\bibinfo{title}{Due to heating of the electrons by the probe beam for
  temperatures less than 100 {eV}, these measurements are most likely an
  overestimate of the true electron temperature.}

\bibitem{bondarenko_collisionless_2017}
\bibinfo{author}{Bondarenko, A.~S.} \emph{et~al.}
\newblock \bibinfo{title}{Collisionless momentum transfer in space and
  astrophysical explosions}.
\newblock \emph{\bibinfo{journal}{Nat Phys}} \textbf{\bibinfo{volume}{13}},
  \bibinfo{pages}{573--577} (\bibinfo{year}{2017}).

\bibitem{germaschewski_the-plasma_2016}
\bibinfo{author}{Germaschewski, K.} \emph{et~al.}
\newblock \bibinfo{title}{The plasma simulation code: A modern particle-in-cell
  code with patch-based load-balancing}.
\newblock \emph{\bibinfo{journal}{J. Comput. Phys.}}
  \textbf{\bibinfo{volume}{318}}, \bibinfo{pages}{305 -- 326}
  (\bibinfo{year}{2016}).

\bibitem{fox_kinetic_2018}
\bibinfo{author}{Fox, W.} \emph{et~al.}
\newblock \bibinfo{title}{Kinetic simulation of magnetic field generation and
  collisionless shock formation in expanding laboratory plasmas}.
\newblock \emph{\bibinfo{journal}{Phys. Plasmas}}
  \textbf{\bibinfo{volume}{25}}, \bibinfo{pages}{102106}
  (\bibinfo{year}{2018}).

\bibitem{wright_early-time_1971}
\bibinfo{author}{Wright, T.~P.}
\newblock \bibinfo{title}{Early-time model of laser plasma expansion}.
\newblock \emph{\bibinfo{journal}{Physics of Fluids}}
  \textbf{\bibinfo{volume}{14}}, \bibinfo{pages}{1905} (\bibinfo{year}{1971}).

\end{thebibliography}


\end{document}